\documentclass[]{article}

\usepackage{graphicx}
\usepackage{color}
\usepackage{subfigure}
\usepackage{float}
\usepackage{authblk}
\usepackage[left=2cm,right=2cm,top=2cm,bottom=2cm]{geometry}

\begin{document}

\title{Multiphoton processes by conditional measurements in the atom-field interaction}
\author[1]{Jorge A. Anaya-Contreras}
\author[1]{Arturo Z\'u\~niga-Segundo}
\author[2]{Francisco Soto-Eguibar}
\author[2]{H. Moya-Cessa}
\affil[1]{Instituto Polit\'ecnico Nacional. ESFM, Departamento de F\'isica. Edificio 9, Unidad Profesional “Adolfo L\'opez Mateos, CP 07738 Ciudad de M\'exico, M\'exico}
\affil[2]{Instituto Nacional de Astrof\'{\i}sica, \'Optica y Electr\'onica, INAOE. Luis Enrique Erro 1, Santa Mar\'ia Tonantzintla, Puebla, 72840 M\'exico}

\maketitle

\begin{abstract}
We show that it is possible to add or subtract many photons from a cavity field by interacting it resonantly with a two-level atom. The atom, after entangling with the field inside the cavity and exiting it, may be measured in one of the Schmidt states,  producing a multiphoton process (in the sense that can add or annihilate more photons than a single transition allows), i.e., adding or subtracting several photons from the cavity field. 
\end{abstract}

\section{Introduction}
Conditional measurements in the atom field interaction \cite{Kurizki1,Vogel,Garraway,Kurizki2,Foster}, i.e. projections of the entangled state  onto a selected atomic state once the   atom and field have interacted and the atom has exit the cavity, carrying the information of the field inside it, have been used to produce nonclassical states of light such as Schr\"odinger cat states \cite{Kurizki1} or more complex superpositions of coherent states \cite{Vogel,Garraway,Kurizki2} that may lead to generation of numbers states.   

Gosh and Gerry \cite{Gosh} have shown that a class of nonclassical states may be produced by such conditional measurements which Luis \cite{Luis} called paradoxical, as he showed that a measurement leading to absorption of light produced an increase of the number of photons. More recently, Valverde and Baseia \cite{Baseia} have studied such paradoxical generation of nonclassical states in models of interpolating Hamiltonians, in particular of the (nonlinear) Jaynes-Cummings type.

Here we  show that conditional measurements can lead efficiently to multiphoton processes \cite{Dantsker}, and in this way, a resonant  interaction that allows single-photon exchange of energy between atom and field \cite{Shore} may generate, via highly probable conditional measurements, the creation or annihilation  of more than one photons in the cavity. 

Because the atom-field interaction may be modeled in other type of systems, namely ion-laser interactions \cite{Wall97,segundo}, the results shown here may be translated to such systems, where one could use this effect to cool trapped ions.

\section{Conditional measurements}
Consider the interaction Hamiltonian for the on-resonance interaction of an
atom and a quantized field ($\hbar$ is set to one)
\begin{eqnarray}
H_I=\lambda(\hat{a}\sigma_+ + \hat{a}^{\dagger}\sigma_-),
\end{eqnarray}
the operators $\hat{a}$ and $\hat{a}^{\dagger}$ are the annihilation and creation operators, the parameter $\lambda$ is the so-called Rabi frequency and the atomic operators, $\sigma_-$ and $\sigma_+$, are the lowering and raising Pauli matrices, respectively. We can obtain the evolution operator, $\hat{U}_I(t)=e^{-i\hat{H}_It}$, by expanding the exponential in Taylor series; the even and odd powers may be written as (we use the Pauli matrices in their $2\times 2$ matrix notation) \cite{Stenholm}
\begin{eqnarray}
\hat{H}_I^{2k}=\lambda^{2k}
 \left(
\begin{array}{ll}
(\sqrt{\hat{a}\hat{a}^{\dagger}})^{2k} &0 \\
0  & (\sqrt{\hat{a}^{\dagger}\hat{a}})^{2k}
\end{array}
\right),
\end{eqnarray}
and
\begin{eqnarray}
\hat{H}_I^{2k+1}=\lambda^{2k+1}
 \left(
\begin{array}{ll}
0 &(\sqrt{\hat{a}\hat{a}^{\dagger}})^{2k+1} \hat{V}\\
\hat{V}^{\dagger}(\sqrt{\hat{a}\hat{a}^{\dagger}})^{2k+1}  & 0
\end{array}
\right),
\end{eqnarray}
with $\hat{V}=\frac{1}{\sqrt{\hat{a}\hat{a}^{\dagger}}}\hat{a}$ the London phase operator \cite{London,Susskind}.
So, we may obtain the evolution operator in the form
\begin{eqnarray}
\hat{U}_I(t)=
 \left(
\begin{array}{ll}
\cos\left(  \lambda t\sqrt{\hat{a}\hat{a}^{\dagger}}\right)  &-i\sin\left(  \lambda t \sqrt{\hat{a}\hat{a}^{\dagger}}\right)  \hat{V}\\
-i\hat{V}^{\dagger}\sin\left(  \lambda t\sqrt{\hat{a}\hat{a}^{\dagger}} \right)  & \cos\left(  \lambda t\sqrt{\hat{a}^{\dagger}\hat{a}}\right) 
\end{array}
\right).
\end{eqnarray}
If we consider an initial
coherent state for the field, $|\alpha\rangle$, with its distribution in terms of number states given by
\begin{eqnarray}
|\alpha\rangle=e^{-\frac{|\alpha|^2}{2}}\sum_{n=0}^{\infty} \frac{\alpha^n}{\sqrt{n!}}|n\rangle,
\end{eqnarray}
where $|n\rangle$ are the Fock states, and an excited state for the atom, the initial wavefunction is given by
\begin{eqnarray}
|\psi(0)\rangle=
 \left(
\begin{array}{l}
|\alpha\rangle\\
0  
\end{array}
\right),
\end{eqnarray}
we obtain the evolved wavefunction as $|\psi(t)\rangle=\hat{U}_I(t)|\psi(0)\rangle$, that in the (complete) ket notation is written as
\begin{eqnarray}
|\psi(t)\rangle=|c\rangle|e\rangle+|s\rangle|g\rangle,
\end{eqnarray}
where the unnormalized wavefunctions $|c\rangle$ and $|s\rangle$ are given by
\begin{eqnarray}
|c\rangle=\cos\left(  \lambda t\sqrt{\hat{a}\hat{a}^{\dagger}}\right) |\alpha\rangle, \qquad
|s\rangle=-iV^{\dagger}\sin\left(  \lambda t \sqrt{\hat{a}\hat{a}^{\dagger}}\right) |\alpha\rangle.
\end{eqnarray}
We now write the above wavefunction in the rotated atomic basis
\begin{eqnarray}
|+\rangle=\cos\theta|e\rangle+e^{-i\phi}\sin\theta|g\rangle, \qquad |-\rangle=e^{i\phi}\sin\theta|e\rangle-\cos\theta|g\rangle ,
\end{eqnarray}
such that $|\psi\rangle$ may be written as a Schmidt decomposition \cite{Gerry}
\begin{eqnarray}
|\psi\rangle=\sqrt{\lambda_+}|+\rangle|\psi_+\rangle+\sqrt{\lambda_-}|-\rangle|\psi_-\rangle ,
\end{eqnarray}
where $ \lambda_{\pm}=\frac{1}{2}\pm \sqrt{W^2+r^2}  $ are the eigenvalues of the atomic density matrix an the orthonormal field wavefunctions are written as
\begin{eqnarray}
|\psi_+\rangle=\frac{1}{\sqrt{\lambda_+}}\left(\cos\theta
|c\rangle+e^{i\phi}\sin\theta|s\rangle\right), \qquad
|\psi_-\rangle=\frac{1}{\sqrt{\lambda_-}}\left(\sin\theta
e^{-i\phi}|c\rangle-\cos\theta|s\rangle\right).
\end{eqnarray}
The quantities $r$, $W$, $\phi$ and $\theta$ are related as follows
\begin{eqnarray}
\langle c|s\rangle =r e^{-i\phi}, \qquad 2W=\langle c|c\rangle - \langle s|s\rangle,
\end{eqnarray}
and
\begin{eqnarray}
\theta =\frac{1}{2}\tan^{-1}\frac{r}{W}.
\end{eqnarray}
By measuring the atom as it exits the cavity in the $|+\rangle$ state, i.e. doing a conditional measurement of the atom in that specific Schmidt state, the field is projected to the normalized state $|\psi_+\rangle$. From it, we can calculate the average number of photons, $N=\langle \psi_+|a^{\dagger}a|\psi_+\rangle$. We plot this quantity as a function of $\lambda t$ in Figure 1, where it is shown that an increase (after a severe decrease) of the average number of photons would occur for certain times (where the maximums happens).  Of course, although it may be seen as a  surprising effect such nonlinear generation of photons in the cavity, it may be understood by the fact that a distribution of photons is initially considered as well as the fact that the conditional measurement is in itself a non-unitary process. 
\begin{figure}[H]
	\begin{center}
		\includegraphics[scale=0.3]{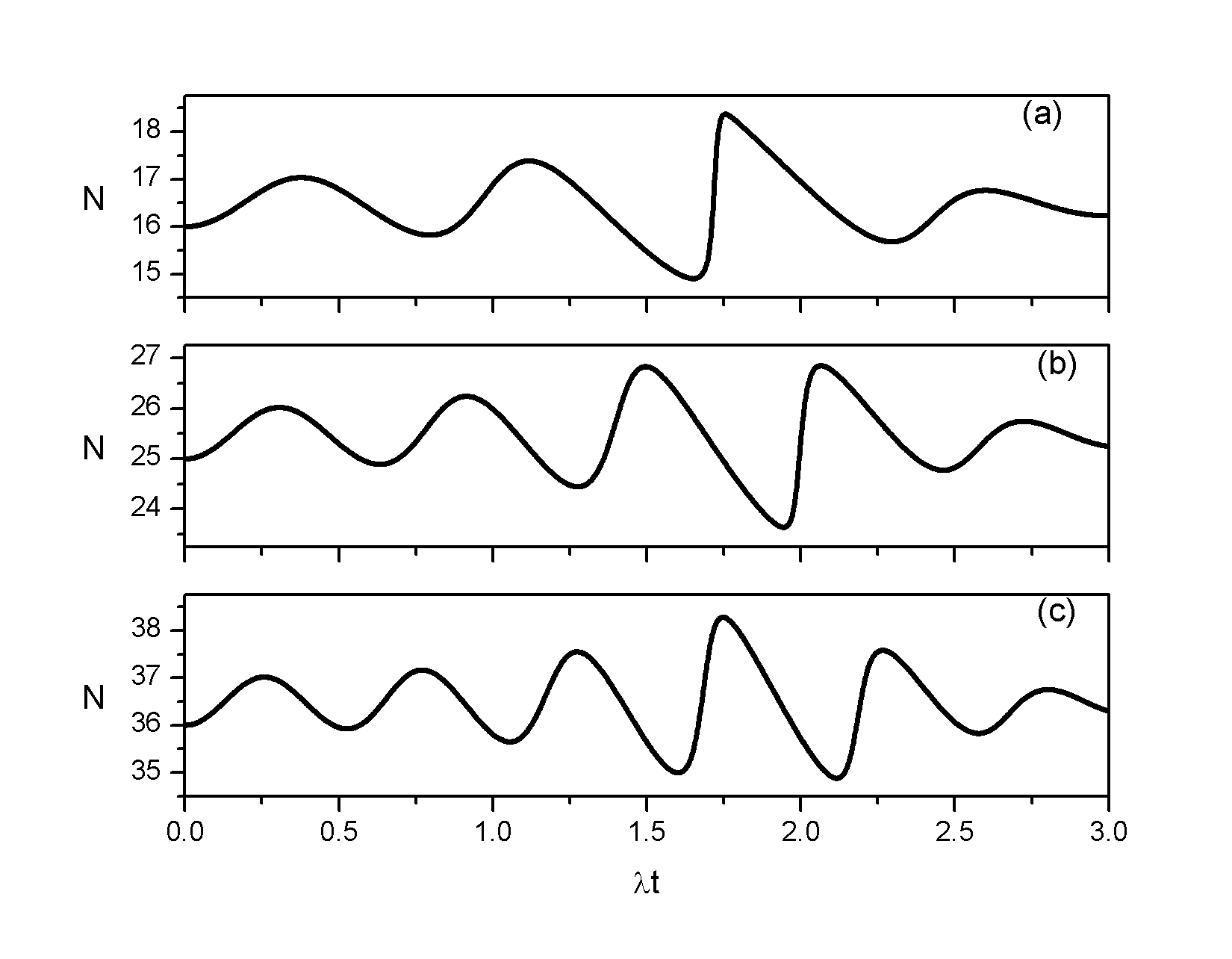}
		\caption{Average number of photons, $N=\langle \psi_+|\hat{a}^{\dagger}\hat{a}|\psi_+\rangle$. The field is initially in a coherent state with (a) $\alpha=4$, (b) $\alpha=5$ and (c) $\alpha=6$.}\label{fig1}
	\end{center}
\end{figure}
In Figure 2, we plot (dashed line) the probability of such events occurring. We see that for the first maximum of the solid line in Figure 1, $\lambda t\approx 1.75$, the probability shown in Figure 2 is a minimum yielding a probability of approximately $3$ percent. It may be also seen that all the maximums happen with minimum probability, although this probability increases such that the second relevant \textquotedblleft multiphoton process\textquotedblright has a probability to occur greater than 20 percent.
\begin{figure}[H]
	\begin{center}
		\includegraphics[scale=0.3]{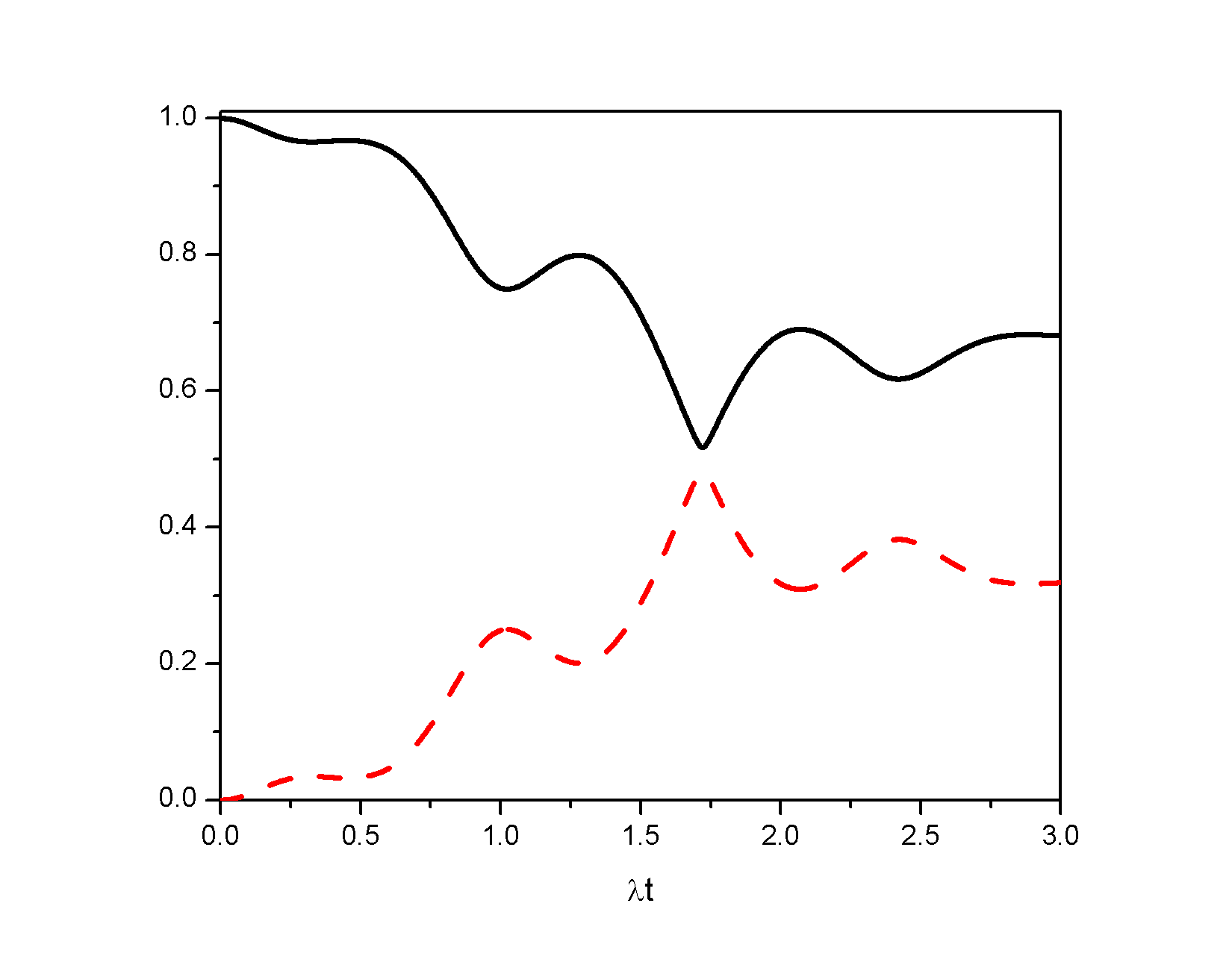}
		\caption{Probabilities (as  functions of $\lambda t$) of measuring the atom after it exits the cavity in the ground state (solid line) or excited state (dash line).}\label{fig2}
	\end{center}
\end{figure}
In Figure 3, we plot the Mandel-$Q$ parameter \cite{Mandel}
\begin{eqnarray}
Q=\frac{\langle \psi_+|(\hat{a}^{\dagger}\hat{a})^2|\psi_+\rangle-\langle \psi_+|\hat{a}^{\dagger}\hat{a}|\psi_+\rangle^2}{\langle \psi_+|\hat{a}^{\dagger}\hat{a}|\psi_+\rangle}-1
\end{eqnarray}
as a function of $\lambda t$ where the generation of nonclassical states may be seen for times for which the Mandel-$Q$ parameter is less than zero. 
\begin{figure}[H]
	\begin{center}
		\includegraphics[scale=0.3]{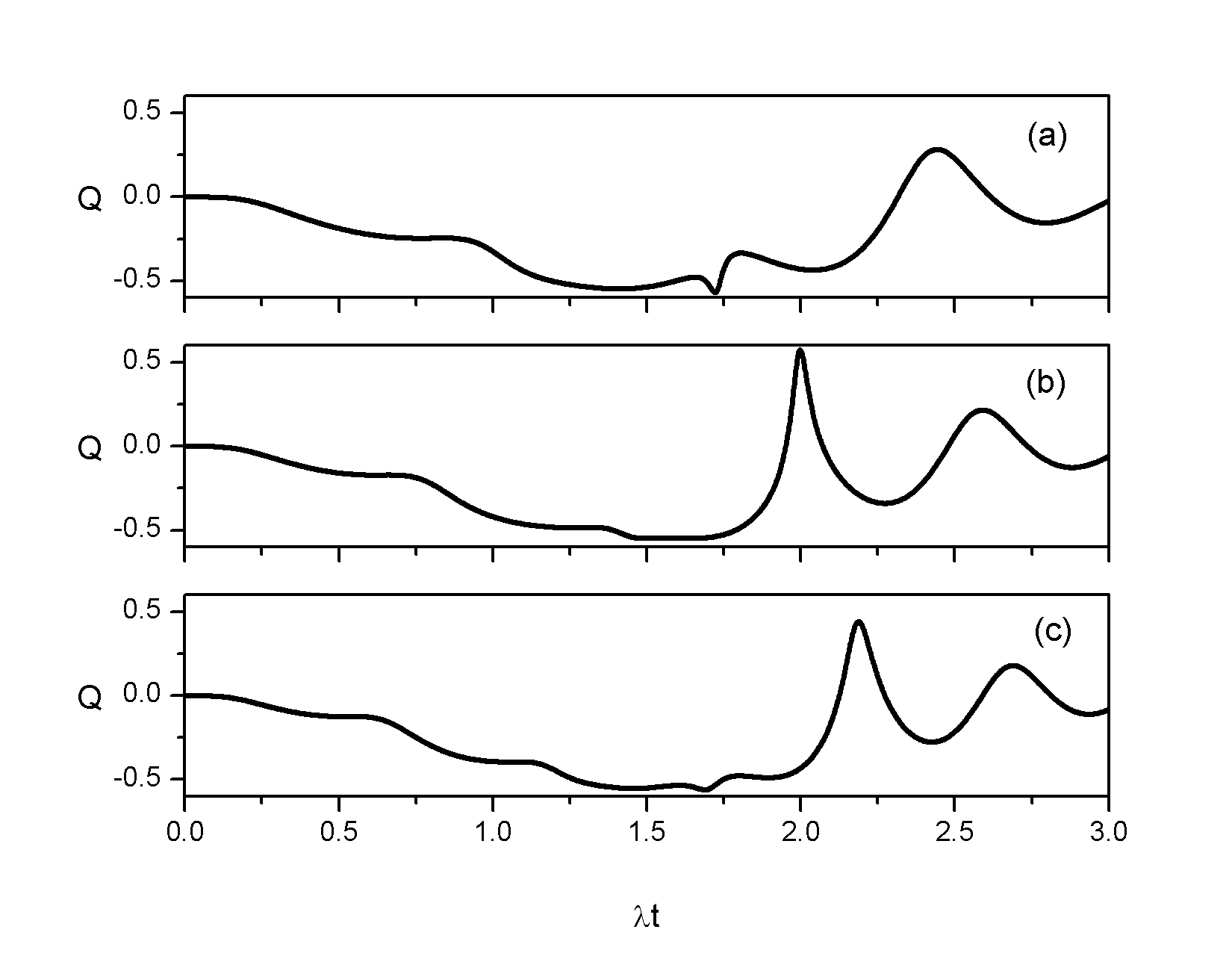}
		\caption{Mandel-$Q$ parameter as a function of $\lambda t$ for the state $|\psi_+\rangle$ the field is initially in a coherent state with (a) $\alpha=4$, (b) $\alpha=5$ and (c) $\alpha=6$.}\label{fig3}
	\end{center}
\end{figure}
In Figure 4 (b), we plot the photon distribution, $P_n=|\langle \psi_+|n \rangle|^2$ as a function of $n$ for $\alpha=4$ and $\lambda t\approx 1.75$. A deviation from the original distribution, Figure 4 (a), centered at $n=16$ may be seen, as the one obtained from a conditional measurement may be seen centered at about $n= 20$. Therefore, an increase of the average number of photons greater than one, i.e. a multiphoton process, takes place. This figure resembles that of a squeezed state \cite{Loudon}. In fact, squeezed light has been predicted in the Jaynes-Cummings model by Kuklinski and  Madajczyk \cite{Kukl} and its interaction with two-level atoms studied by Satyanarayana {\it et al.} \cite{Satya} and Moya-Cessa and Vidiella-Barranco \cite{squeezed}.
\begin{figure}[H]
	\begin{center}
		\includegraphics[scale=0.4]{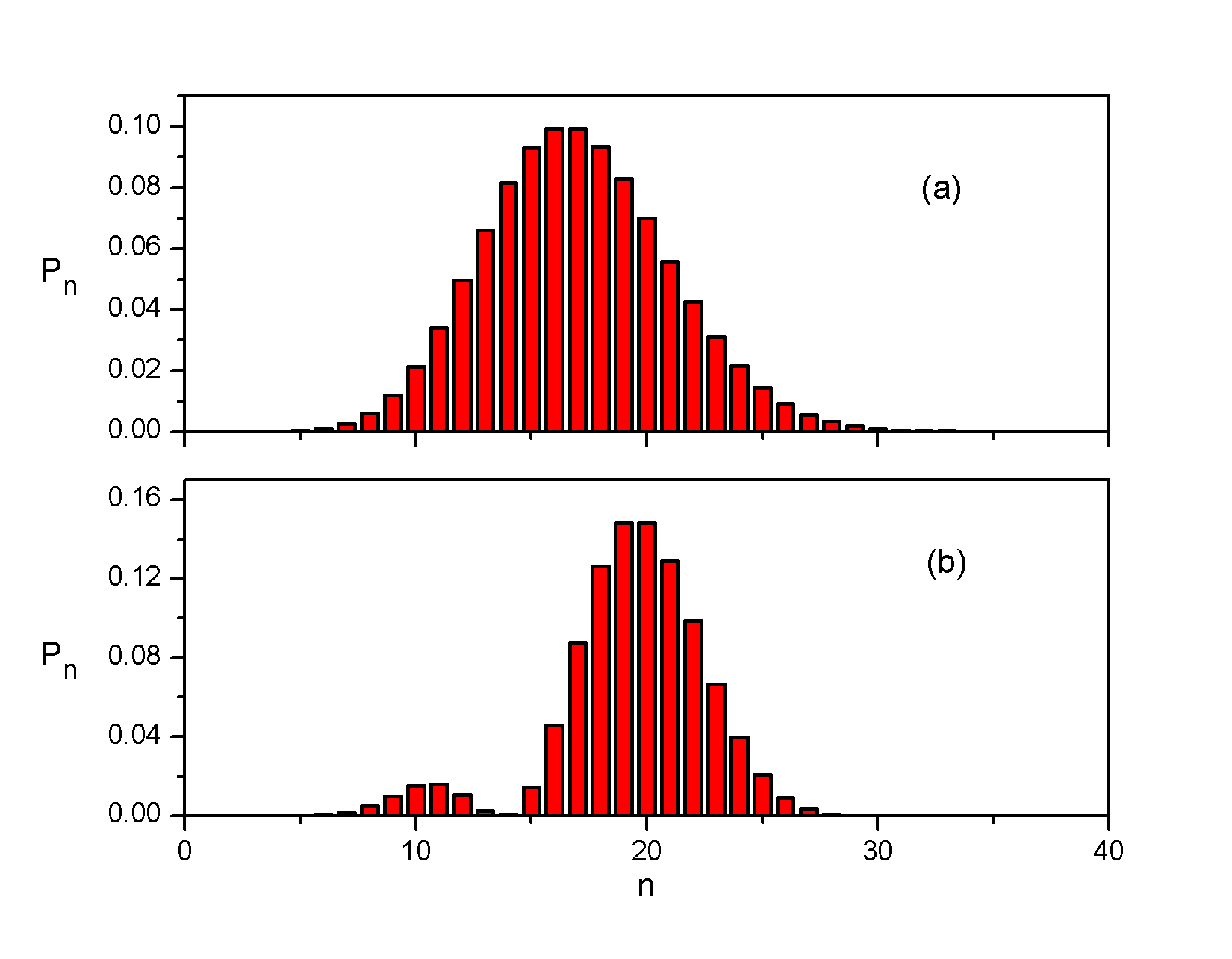}
		\caption{Photon distribution for (a) a coherent state with $\alpha=4$ and (b) the conditional state $|\psi_+\rangle$, with  $\alpha=4$ and an interaction time $t\approx 1.75/\lambda$. }\label{fig4}
	\end{center}
\end{figure}
\begin{figure}[h!]
\begin{center}
\includegraphics[scale=0.4]{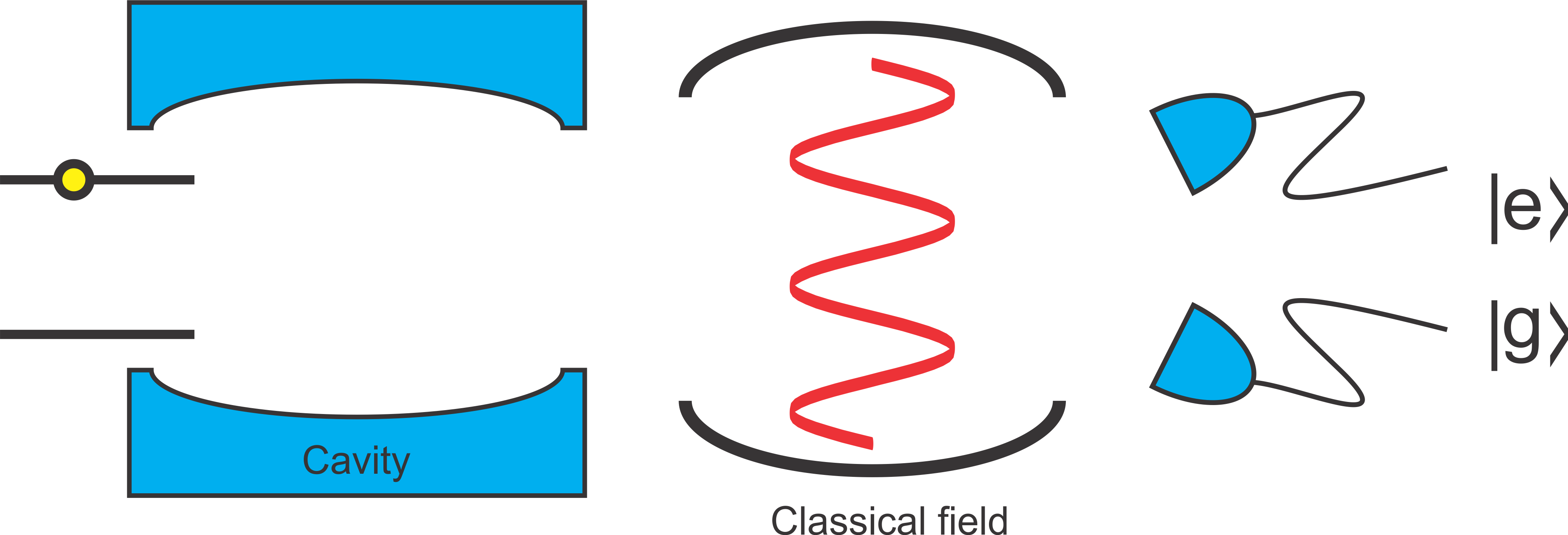}
\caption{Possible setup for the measurement of Schmidt states. The atom is initially prepared in an excited state, interacts with a cavity field which leaves then entangled. The atom is then properly rotated by a classical field and   a further measurement in the excited or ground state produces a complete measurement of the $|+\rangle$ or $|-\rangle$ state,  respectively.}\label{fig5}
\end{center}
\end{figure}

\section{Conclusions}
We have shown that, after an atom  resonantly interacts with a field, the average number of photons inside the cavity is severely affected by measuring it when it exits it. We call this a multiphoton process as the atom-field interaction allows the exchange of only  single photons, but the conditional measurement generates the creation or absorption of several photons in the cavity. We have shown, numerically, that such effects are related to the initial number of photons in the cavity for both coherent states and thermal distributions giving rise to the addition or subtraction proportional to the square of the initial average number of photons.

\end{document}